\documentstyle[12pt,axodraw,epsf]{article} 
\catcode`@=11
\newcount\@tempcntc
\def\@citex[#1]#2{\if@filesw\immediate\write\@auxout{\string\citation{#2}}\fi
\@tempcnta\z@\@tempcntb\m@ne\def\@citea{}\@cite{\@for\@citeb:=#2\do
{\@ifundefined
{b@\@citeb}{\@citeo\@tempcntb\m@ne\@citea\def\@citea{,}{\bf
?}\@warning
       {Citation `\@citeb' on page \thepage \space undefined}}%
    {\setbox\z@\hbox{\global\@tempcntc0\csname b@\@citeb\endcsname\relax}%
     \ifnum\@tempcntc=\z@ \@citeo\@tempcntb\m@ne
       \@citea\def\@citea{,}\hbox{\csname b@\@citeb\endcsname}%
     \else \advance\@tempcntb\@ne \ifnum\@tempcntb=\@tempcntc
      \else\advance\@tempcntb\m@ne\@citeo
      \@tempcnta\@tempcntc\@tempcntb\@tempcntc\fi\fi}}\@citeo}{#1}}
\def\@citeo{\ifnum\@tempcnta>\@tempcntb\else\@citea\def\@citea{,}%
  \ifnum\@tempcnta=\@tempcntb\the\@tempcnta\else
   {\advance\@tempcnta\@ne\ifnum\@tempcnta=\@tempcntb \else
   \def\@citea{--}\fi
   \advance\@tempcnta\m@ne\the\@tempcnta\@citea\the\@tempcntb}\fi\fi}
   \catcode`@=12

\begin{document}

\begin{flushright}
MPI/PhT/98--38\\ 
hep-ph/9805373\\ 
April 1998
\end{flushright}

\begin{center}
{\LARGE {\bf Higgs Scalar-Pseudoscalar Mixing in the}}\\[0.4cm] 
{\LARGE {\bf Minimal Supersymmetric Standard Model }}\\[2.4cm]
{\large Apostolos Pilaftsis}\footnote[1]{E-mail address:
  pilaftsi@mpmmu.mpg.de}\\[0.4cm] 
{\em Max-Planck-Institut f\"ur Physik, F\"ohringer Ring 6, 80805 
                                                 Munich, Germany}
\end{center}
\vskip1.7cm \centerline{\bf  ABSTRACT}  
In  the minimal supersymmetric   Standard Model, the  heaviest CP-even
Higgs  boson $H$ and the  CP-odd Higgs scalar  $A$ are predicted to be
almost  degenerate in mass at the  tree  level, for the wide kinematic
range $M_A    >  2M_Z$ and  $\tan\beta \ge     2$.   However, if large
soft-CP-violating  Yukawa interactions involving  scalar quarks of the
third generation are present in the theory,  then the CP invariance of
the Higgs     potential  can be maximally    broken  beyond   the Born
approximation, and the high degree of mass  degeneracy between $H$ and
$A$  may be lifted through a  sizeable $HA$  mixing.  After taking the
CP-odd tadpole renormalization of the $A$  boson into account, we find
that the small mass difference $M_H-M_A$,  which is about $1$\% of the
$A$-boson mass at the tree level, can be substantially enhanced to the
25\% level at the one-loop order.  We also  find that the loop-induced
mixing between the lightest CP-even Higgs boson  $h$ and the $A$ boson
may be  of comparable  size to $M_h$.    We briefly  discuss the  main
phenomenological implications of the predicted  $hA$ and $HA$  mixings 
for the general Higgs-boson mass spectrum and for CP-violating
observables at collider and lower energies.

\newpage

It  is  known  \cite{GHKD} that  the Higgs   potential of  the minimal
supersymmetric  Standard  Model    (MSSM)   is invariant   under   the
transformations  of charge conjugation and  parity   (CP) at the  tree
level.  The reason is  that supersymmetry (SUSY) imposes an additional
(holomorphic) symmetry  on the  Higgs  sector of a   general two-Higgs
doublet  model,  which entails   flavour  conservation in   tree-level
neutral   currents  and  absence  of  CP-violating scalar-pseudoscalar
mixings  in the Born   approximation.  Beyond the Born  approximation,
recent studies have shown  that CP invariance  of the Higgs  potential
may in principle be broken spontaneously through radiative corrections
\cite{NM}  if   the CP-odd  Higgs  scalar  $A$   is sufficiently light
\cite{APNH,KL}.  However,  this possibility has  now been ruled out by
experiment \cite{APNH,KL}.

Here, we shall study in more detail another interesting possibility of
CP non-conservation within the context of  the MSSM.  In this case, CP
violation in the  Higgs sector is induced by  loop effects due to  the
presence of additional interactions  in  the theory, which violate  CP
explicitly.  For example, such  CP-violating interactions may occur in
the trilinear Yukawa  couplings of the  Higgs fields to scalar quarks. 
However, the phenomenological viability of such a scenario of explicit
CP  violation is  often  questioned in the  literature.   The standard
reasoning for the  latter goes as follows.   Unless  the scalar quarks
have TeV masses, all new  CP-violating phases in  the MSSM not present
in the SM \cite{EFN,DGH,KO} must be suppressed at least by a factor of
order $10^{-3}$, otherwise  one-loop  effects may exceed  the  current
experimental  limit on the  neutron   electric dipole  moment (EDM).   
According to these general arguments, CP violation in the Higgs sector
of the MSSM, which arises  at one loop,  was estimated to be  dismally
small, so as  to bear any  phenomenological relevance \cite{APNH}.  In
this paper, we shall show that without  any further assumptions on the
model, this rough estimate  based  on the neutron   EDM limit and  the
naive counting  of loop suppression  factors is  fairly inaccurate and
may   be  misleading   in  general.  In    particular,  we  find  that
scalar-pseudoscalar mixings can still be large in the MSSM giving rise
to observable  CP-violating phenomena, if  the Yukawa sector involving
the   scalar top     and  bottom quarks   contains  relatively   large
CP-violating  couplings compatible  with universal boundary conditions
imposed by minimal supergravity models at the unification scale $M_X$.

The MSSM introduces several new phases  in the theory which are absent
in  the SM \cite{EFN}.  The  mixing mass parameter $\mu$ involving the
two Higgs      chiral    superfields in the       superpotential,  the
soft-SUSY-breaking  gaugino   masses  $m_\lambda$ (with   $\lambda   =
\tilde{g},\ \tilde{W}$  and  $\tilde{B}$ representing  the gauginos of
the SU(3)$_c$, SU(2)$_L$ and U(1)$_Y$ gauge groups, respectively), the
soft bilinear  Higgs  mixing  mass $m^2_{12}$ (sometimes   denoted  as
$B\mu$ in the  literature) and  the  soft trilinear Yukawa   couplings
$A_f$ are all complex numbers.  If the universality condition at $M_X$
is assumed,  the gaugino masses  $m_\lambda$ are then related  to each
other and have the same phase, while the different trilinear couplings
$A_f$ are  all equal, {\em  i.e.}, $A_f = A$.  Not   all phases of the
four complex parameters $\{  \mu ,\ m^2_{12},\ m_\lambda  ,\ A \}$ are
independent  of the fields  \cite{DGH}.   For instance,  one  can make
$m_\lambda$ real by redefining the  gaugino field $\lambda$.  Also, as
we will see below, minimization conditions on the Higgs potential lead
to the constraint that the phase of  $m^2_{12}$ should be equal to the
phase difference of the two Higgs doublets in the  MSSM.  As a result,
there are only two independent CP-violating phases in this constrained
version of  the MSSM.  Usually, these  are taken to be  arg($\mu$) and
arg($A$).  Limits coming from  the  electron EDM  may be avoided  to a
great extent by requiring that the  phases of $m_\lambda$ and $\mu$ be
aligned  to $m^2_{12}$,  {\em i.e.},  $\Im  m  (m^*_\lambda \mu)  = 0$
\cite{KO,FOS}.  In this  case, chargino  and neutralino mass  matrices
are real and conserve CP.  Furthermore,  it has been argued \cite{FOS}
that  bounds obtained from  the neutron EDM leave arg($A$) essentially
unconstrained,  even for slightly smaller than  TeV soft scalar masses
\cite{FOS}.  Notice that constraints on   the scalar- top and   bottom
sector do not come directly from the neutron EDM but rather indirectly
via the universality conditions  at $M_X$.  From the above discussion,
it is clear that for soft scalar masses in the range  $0.5 \le M_0 \le
1$ TeV, arg($A$)   can  safely be considered  to   be the only   large
CP-violating phase in  the theory of  order unity.  As  a consequence,
significant scalar-pseudoscalar mixings in  the MSSM are only expected
to come from loop  effects of scalar   top and bottom  quarks, whereas
chargino   and  neutralino   contributions being proportional   to the
vanishingly small CP-violating phase arg($\mu$) may be neglected.

We  start  our discussion by  considering the   Higgs potential of the
MSSM. Because of the holomorphic property of SUSY, one needs two Higgs
doublets at  least,   denoted as $\tilde{\Phi}_1$ and   $\Phi_2$, with
opposite hypercharges, $Y(\Phi_2) = - Y(\tilde{\Phi}_1) = 1$, in order
to give masses to both  up- and down-  quark families and, at the same
time,   cancel the triangle  anomalies  induced by  the fermionic SUSY
partners  of the   Higgs     field.   After  integrating    over   the
Grassmann-valued  coordinates in the   SUSY  action and including  the
soft-SUSY-breaking  masses for the Higgs  fields,  one arrives at  the
Lagrangian describing the Higgs potential
\begin{eqnarray}
  \label{LV}
{\cal L}_V &=& \mu^2_1 (\Phi_1^\dagger\Phi_1)\, +\, 
\mu^2_2 (\Phi_2^\dagger\Phi_2)\, +\, m^2_{12} (\Phi_1^\dagger \Phi_2)\, 
+\, m^{*2}_{12} (\Phi_2^\dagger \Phi_1)\, +\, 
\lambda_1 (\Phi_1^\dagger \Phi_1)^2\nonumber\\
&&+\, \lambda_2 (\Phi_2^\dagger \Phi_2)^2\, +\, 
\lambda_3 (\Phi_1^\dagger \Phi_1)(\Phi_2^\dagger \Phi_2)\, +\, 
\lambda_4 (\Phi_1^\dagger \Phi_2)(\Phi_2^\dagger \Phi_1)\, ,
\end{eqnarray}
where  $\Phi_1 =   -i\tau_2 \tilde{\Phi}_1^*$ ($\tau_2$  is the  Pauli
matrix) and
\begin{eqnarray}
  \label{LVpar}
\mu^2_1 &=& -m^2_1 - |\mu|^2\, ,\qquad \mu^2_2\ =\ -m^2_2 - |\mu|^2\, ,
                                                            \nonumber\\
\lambda_1 &=& \lambda_2\ =\ -\, \frac{1}{8}\, (g^2 + g'^2)\, ,\qquad
\lambda_3\ =\ -\frac{1}{4}\, (g^2 -g'^2)\, ,\qquad 
\lambda_4\ =\ \frac{1}{2}\, g^2\, .
\end{eqnarray}
The complex  parameter $m^2_{12}$  in  Eq.\ (\ref{LV}) as well  as the
real  parameters  $m^2_1$ and  $m^2_2$ in  Eq.\ (\ref{LVpar}) are soft
Higgs-scalar   masses.  Furthermore,   $g$  and  $g'$  are   the usual
SU(2)$_L$ and     U(1)$_Y$  gauge  couplings, respectively.     It  is
interesting  to remark that  the  MSSM  Higgs  potential ${\cal  L}_V$
contains fewer  quartic couplings than  that  of the general two-Higgs
doublet model; all quartic  couplings $\lambda_i$ in ${\cal  L}_V$ are
uniquely specified by  SUSY.  This makes the  Higgs sector of the MSSM
highly predictive.  

In order to determine the ground state of the MSSM Higgs potential, we
consider the linear decompositions of the Higgs fields
\begin{equation}
  \label{Phi12}
\Phi_1\ =\ \left( \begin{array}{c}
\phi^+_1 \\ \frac{1}{\sqrt{2}}\, ( v_1\, +\, H_1\, +\, iA_1)
\end{array} \right)\, ,\qquad
\Phi_2\ =\ e^{i\xi}\, \left( \begin{array}{c}
\phi^+_2 \\  \frac{1}{\sqrt{2}}\, ( v_2 \, +\, H_2\, +\, iA_2 )
 \end{array} \right)\, ,
\end{equation}
where $v_1$ and $v_2$ are the  moduli of the vacuum expectation values
(VEV's) of the  Higgs doublets and $\xi$ is  their relative phase. The
positive parameters $v_1$ and $v_2$  and the phase $\xi$ are  entirely
fixed by  the minimization conditions  on  ${\cal L}_V$.  This  can be
accomplished by  requiring  the  vanishing of   the  following tadpole
parameters:
\begin{eqnarray}
  \label{TH1}
T_{H_1} &\equiv& \langle\,\frac{\partial {\cal L}_V}{\partial H_1}\,\rangle
\ =\ 
v_1\, \Big[\, \mu^2_1\ +\ \Re e (m^2_{12}e^{i\xi})\, \tan\beta\ -\, 
\frac{1}{2}\, M^2_Z\, \cos 2\beta\, \Big]\, ,\\
  \label{TH2}
T_{H_2} &\equiv& \langle\,\frac{\partial {\cal L}_V}{\partial H_2}\,\rangle
\ =\ 
v_2\, \Big[\, \mu^2_2\ +\ \Re e (m^2_{12} e^{i\xi})\, \cot\beta\ 
+\, \frac{1}{2}\, M^2_Z\, \cos 2\beta\, \Big]\, ,\\
  \label{TA1}
T_{A_1} &\equiv& \langle\,\frac{\partial {\cal L}_V}{\partial A_1}\,\rangle
\ =\ v_2 \Im m (m^2_{12} e^{i\xi})\, ,\\
  \label{TA2}
T_{A_2} &\equiv& \langle\,\frac{\partial {\cal L}_V}{\partial A_2}\,\rangle
\ =\ -v_1 \Im m (m^2_{12} e^{i\xi})\, ,
\end{eqnarray}
where $\tan\beta = v_2/v_1$  and $M^2_Z = (g^2  + g'^2) v^2/ 4$ is the
$Z$-boson  mass  squared with $v^2  =  v^2_1  + v^2_2$.  Variations of
${\cal L}_V$    with   respect to $\phi^+_1$   and   $\phi^+_2$ vanish
identically, reflecting the fact that  a physical ground state  should
conserve charge \cite{VRS}. If we now perform the orthogonal rotation
\begin{equation}
  \label{G0A}
\left( \begin{array}{c} A_1 \\ A_2\end{array}\right)\ =\
\left( \begin{array}{cc} \cos\beta & -\sin\beta \\ 
\sin\beta & \cos\beta\end{array}\right)\,
\left( \begin{array}{c} G^0 \\ A\end{array}\right)\, ,
\end{equation}
the Higgs  potential  shows up a   flat direction with respect to  the
$G^0$ field, {\em  i.e.},  $\langle \partial  {\cal  L}_V/\partial G^0
\rangle  =  0$.   Then, the $G^0$  field   becomes  the true  would-be
Goldstone boson eaten by the longitudinal component of  the $Z$ boson. 
In this  weak basis, the tree-level mass  matrix of the CP-odd scalars
becomes diag$(0,\ M^2_A)$, where $M^2_A =  \Re e (m^2_{12} e^{i\xi}) /
(s_\beta c_\beta)$  is  the   tree-level   $A$-boson mass   squared.   
Furthermore, the   orthogonal    rotation (\ref{G0A})   leads   to   a
non-trivial CP-odd tadpole parameter given by
\begin{equation}
  \label{TadA}
T_A\ \equiv \ \langle\, \frac{\partial {\cal L}_V}{\partial A}\, \rangle\ =\
        -\, v\, \Im m ( m^2_{12} e^{i\xi} )\, .
\end{equation}
As   has  been   shown  explicitly    in \cite{CPodd},  the    tadpole
renormalization  of   the $A$  boson is  very   crucial  to render all
$H_1G^0$, $H_2G^0$, $H_1A$ and $H_2A$ mixings ultra-violet (UV) finite
(see also  Fig.\  1). It is now   important to notice  that  the phase
difference $\xi$ between the two Higgs VEV's is  no more arbitrary but
completely  specified by virtue  of  Eq.\ (\ref{TadA}).   At  the tree
level, one  has $T_A=0$   and $m^2_{12}  e^{i\xi}$ is  a  real number.
Beyond the Born  approximation however, $m^2_{12} e^{i\xi}$ acquires a
small imaginary part.  In fact, the non-vanishing tadpole graph of the
$A$-boson  should  be  compensated  by  the tadpole counter-term  (CT)
$T_A$,    such that  the true  ground   state  of the effective  Higgs
potential does not get  shifted.  Without any  loss of  generality, we
can therefore  redefine $\Phi_2$ as $e^{-i{\rm arg}(m^2_{12})}\Phi_2$,
thereby resulting in a weak basis in which  $\xi = 0$ at zeroth order.
Then, $m^2_{12}$ is real at the tree level but effectively receives an
imaginary  part  at higher orders  which is  determined by the tadpole
renormalization condition on $T_A$.  With this simplification, one can
avoid unnecessary $\xi$-dependent phases in the MSSM.

After including all tadpole contributions, the Lagrangian relevant for
the Higgs-boson mass matrix can be cast into the general form
\begin{equation}
  \label{LHmass}
{\cal L}^H_{\rm mass} \ =\ -\, \frac{1}{2}\ \Big( H_1,\ H_2,\ G^0,\ A\,\Big)\,
\left(\begin{array}{cc} {\cal M}^2_S & {\cal M}^2_{SP} \\
                {\cal M}^2_{PS} & \widehat{\cal M}^2_P \end{array} \right)\, 
\left( \begin{array}{c}
H_1 \\ H_2 \\ G^0 \\ A \end{array} \right)\, .
\end{equation}
Employing the  usual  short-hand notations  $s_x   =\sin x$ and  $\cos
x=c_x$, the $2\times 2$ sub-matrices in  Eq.\ (\ref{LHmass}) are given
by
\begin{eqnarray}
  \label{M2S}
{\cal M}^2_S &=&  \left(\begin{array}{cc} 
c^2_\beta M^2_Z\, +\, s^2_\beta M^2_A\, -\, T_{H_1}/v_1 & 
                                    -s_\beta c_\beta (M^2_Z\, +\, M^2_A) \\
-s_\beta c_\beta (M^2_Z\, +\, M^2_A) &
s^2_\beta M^2_Z\, +\, c^2_\beta M^2_A\, -\, T_{H_2}/v_2\end{array}
                                                              \right)\, ,\\ 
  \label{M2Phat}
\widehat{\cal M}^2_P & =&  \left( \begin{array}{cc} 
-\, \frac{\displaystyle c_\beta T_{H_1} + s_\beta
  T_{H_2}}{\displaystyle  v} 
& \frac{\displaystyle s_\beta T_{H_1} - c_\beta T_{H_2}}{\displaystyle v}\\
\frac{\displaystyle s_\beta T_{H_1} - c_\beta T_{H_2}}{\displaystyle v}&
\quad M^2_A\, -\, 
\frac{\displaystyle s_\beta\tan\beta\, T_{H_1} +
c_\beta\cot\beta\, T_{H_2}}{\displaystyle v} 
\end{array} \right)\, ,\\
  \label{M2SP}
{\cal M}^2_{SP} &=& -\, \frac{T_A}{v} \left(\begin{array}{cc} 
s_\beta & c_\beta \\ - c_\beta & s_\beta \end{array} \right)\, 
\end{eqnarray}
and ${\cal     M}^2_{PS}  =  ({\cal  M}^2_{SP})^T$.   In      the Born
approximation,  the CP-even mass eigenstates  $h$ and $H$ are obtained
by diagonalizing ${\cal M}^2_S$ through the orthogonal transformation
\begin{equation}
  \label{RHiggs}
\left( \begin{array}{c} H_1 \\ H_2\end{array}\right)\ =\
\left( 
\begin{array}{cc} \cos\alpha & -\sin\alpha \\ \sin\alpha & \cos\alpha
\end{array}\right)\,
\left( \begin{array}{c} h \\ H\end{array}\right)\, ,
\end{equation}
with
\begin{equation}
  \label{alpha}
\tan (2\alpha)\ =\ \frac{M^2_A\, +\, M^2_Z}{M^2_A\, -\, M^2_Z}\
\tan (2\beta)\, . 
\end{equation}
The tree-level mass eigenvalues of ${\cal M}^2_S$ are then given by 
\begin{equation}
  \label{MhH}
M^2_{h (H)}\ =\ \frac{1}{2}\ \Big[\, M^2_Z + M^2_A\ -(+)\
\sqrt{(M^2_Z+M^2_A)^2 - 4M^2_A M^2_Z \cos^2 2\beta }\ \Big]\, . 
\end{equation}
Notice that for $\tan\beta \ge 2$, one has  $\cos^2 2\beta \approx 1$,
$M_h \approx M_Z$  and $M_H  \approx M_A$,  that is,  $H$ and $A$  are
nearly degenerate in the  MSSM.  However, radiative corrections due to
the   large top  Yukawa  coupling   affect the $h$-boson    mass in  a
significant manner.   In fact, $h$  can be as heavy as  130 GeV and is
heavier than the $Z$ boson for the largest bulk of the parameter space
\cite{Mh}.  On  the other hand,   the high  degree of mass  degeneracy
between $H$  and $A$ generally persists even  beyond the tree level in
the CP-invariant limit of the theory, especially when $M_A > 2M_Z$ and
$\tan\beta  \ge   2$.  In this  kinematic    range, $\tan\beta \approx
\tan\alpha$  and  quantum effects  seem to   affect  equally $M_H$ and
$M_A$, such that the small mass splitting of $H$ and $A$ remains still
valid \cite{KKRW}.  As we  will see however, large CP-violating Yukawa
couplings of scalar  top and bottom quarks   can give rise to  sizable
$HA$ mixings at the one loop level and to a substantial enhancement of
the    mass  difference      $M_H-M_A$.   In  the      calculation  of
scalar-pseudoscalar self-energy transitions, one   has to include  the
relevant CP-violating mass CT's given by ${\cal M}^2_{SP}$ in order to
arrive at UV-safe analytic results.

We shall now discuss the interactions of the neutral Higgs fields with
the scalar top and bottom quarks in the presence of CP-violating terms
\cite{GH}.  The scalar quarks of  the third generation are represented
by
\begin{equation}
  \label{Qtilde}
\tilde{Q}\ =\ \left(\begin{array}{c} \tilde{t}_L \\ 
\tilde{b}_L \end{array}\right)\, ,\qquad \tilde{U}^*\ =\ \tilde{t}_R\, ,
\qquad \tilde{D}^*\ =\ \tilde{b}_R\, ,
\end{equation}
with $Y(\tilde{Q}) = 1/3$, $Y(\tilde{U})  = -4/3$, and $Y(\tilde{D}) =
2/3$.  If we  denote the   neutral  component of  the Higgs   doublets
$\Phi_1$ and $\Phi_2$  by $\phi^0_1$ and   $\phi^0_2$, there are  then
three contributions of  $\phi^0_1$   and  $\phi^0_2$ to   scalar-quark
masses and their respective couplings. The relevant Lagrangian, ${\cal
  L}^0_V$, receives contributions  from the soft-SUSY-breaking sector,
and from the so-called $F$ and $D$ terms (or components) of the chiral
and vector superfields, respectively.  Specifically, we have
\begin{equation}
  \label{LV0}
{\cal L}^0_V \ =\ {\cal L}^0_{\rm soft}\ +\ {\cal L}^0_F\ +\ 
{\cal L}^0_D\, ,
\end{equation}
where
\begin{eqnarray}
  \label{Lsoft}
-{\cal L}^0_{\rm soft} &=& \tilde{M}^2_Q\, ( \tilde{t}^*_L\tilde{t}_L +
\tilde{b}^*_L\tilde{b}_L )\, +\, \tilde{M}^2_t\, \tilde{t}^*_R\tilde{t}_R\, 
+\, \tilde{M}^2_b\, \tilde{b}^*_R\tilde{b}_R\, 
+\, \Big( f_1 A_b\, \phi^{0*}_1 \tilde{b}^*_R \tilde{b}_L\, 
+\, f_2 A_t\, \phi^0_2 \tilde{t}^*_R \tilde{t}_L\, \nonumber\\
&&+\quad \mbox{H.c.}\,\Big)\,,\nonumber\\
  \label{LF}
-{\cal L}^0_F &=& f^2_1\, \phi^{0*}_1\phi^0_1 ( \tilde{b}^*_L\tilde{b}_L\,
+\, \tilde{b}^*_R\tilde{b}_R )\ +\ 
f^2_2\, \phi^{0*}_2\phi^0_2 ( \tilde{t}^*_L\tilde{t}_L\,
+\, \tilde{t}^*_R\tilde{t}_R )\nonumber\\
&& +\, \Big( \mu f_2\, \phi^{0*}_1 \tilde{t}^*_L \tilde{t}_R\, 
+\, \mu f_1\, \phi^0_2 \tilde{b}^*_L \tilde{b}_R\ +\quad \mbox{H.c.}\,
\Big)\, ,\nonumber\\
  \label{LD}
-{\cal L}^0_D &=& \frac{M^2_Z}{v^2}\, \Big( \phi^{0*}_1\phi^0_1 \, -\, 
\phi^{0*}_2\phi^0_2 \Big)\, \Big[ (1-2e_ts^2_w)\, \tilde{t}^*_L
\tilde{t}_L\, +\, 2e_ts^2_w\, \tilde{t}^*_R\tilde{t}_R\, -\, 
(1+2e_b s^2_w)\, \tilde{b}^*_L \tilde{b}_L\nonumber\\ 
&&+\, 2e_bs^2_w\, \tilde{b}^*_R\tilde{b}_R\, \Big]\, ,
\end{eqnarray}
with $s_w =  \sin\theta_w$ being the  weak mixing angle, $e_t  = 2/3$,
$e_b = -1/3$, $f_1 = \sqrt{2} m_b/v_1$ and $f_2  = \sqrt{2} m_t/ v_2$. 
Furthermore,  $\tilde{M}_Q$,    $\tilde{M}_t$  and  $\tilde{M}_b$  are
soft-scalar quark masses and are usually considered to be all equal to
$M_0$ at  the unification scale  $M_X$.  Notice that all  operators of
dimension four  in  ${\cal    L}_V$  and  ${\cal L}^0_V$  satisfy   an
additional global U(1)$_Q$ symmetry with $Q$ charges: $Q(\Phi_1) = 2$,
$Q(\Phi_2)  =  1$,   $Q(\tilde{Q}) =    0$,  $Q(\tilde{U})  = -2$  and
$Q(\tilde{D})  = - 1$; the  bilinear  operator $\Phi^\dagger_1 \Phi_2$
and the  $A$-dependent trilinear terms  break it only softly. In fact,
if the Higgs-mixing term  $\mu$ in the  superpotential is absent,  the
U(1)$_Q$ symmetry   can appropriately be extended  to  the  whole MSSM
Lagrangian.  As has extensively been discussed  in \cite{CPodd} and is
also valid for the case of the MSSM, the simultaneous soft-breaking of
the   symmetries U(1)$_Q$  and  CP     is sufficient  to  assure   the
renormalizability of  Higgs   scalar-pseudoscalar transitions  to  all
orders in perturbation theory.

It  is straightforward to  obtain   the scalar  top  and  bottom  mass
matrices from ${\cal L}^0_V$ which may conveniently be expressed as
\begin{equation}
  \label{Mscalar}
\widetilde{\cal M}^2_q \ =\ \left( \begin{array}{cc}
\tilde{M}^2_Q\, +\, m^2_q\, +\, \cos 2\beta M^2_Z\, ( T^q_z\, -\,
e_q s^2_w ) & m_q (\mu R_q\, +\, A^*_q )\, e^{i\delta_q}\\ 
m_q (\mu^* R_q\, +\, A_q )\, e^{-i\delta_q} &
\tilde{M}^2_q\, +\, m^2_q\, +\, \cos 2\beta M^2_Z\, e_q s^2_w 
\end{array}\right)\, ,
\end{equation}
where $q=t,b$, $T^t_z = 1/2$, $T^b_z = -1/2$,  $R_t = \cot\beta$, $R_b
= \tan\beta$.  The phase $\delta_q$ is determined from the requirement
that the   scalar-quark mass matrix  becomes  positive  definite  by a
judicious re-definition of  the right-handed scalar quark fields, {\em
  i.e.}, $\tilde{q}_R  \to e^{i\delta_q}\, \tilde{q}_R$ with $\delta_q
=  {\rm  arg}(\mu^*  R_q\, +\, A_q  )$.  In  this weak basis,   we can
diagonalize $\widetilde{\cal M}^2_q$ through the orthogonal rotation
\begin{equation}
  \label{Rscalar}
\left( \begin{array}{c} \tilde{q}_L \\ \tilde{q}_R \end{array} \right)
\ =\ \left( \begin{array}{cc} \cos\theta_q & \sin\theta_q \\
          -\sin\theta_q & \cos\theta_q \end{array} \right)\,
\left( \begin{array}{c} \tilde{q}_1 \\ \tilde{q}_2 \end{array}\right)\, ,
\end{equation}
where the rotation angle $\theta_q$ may be obtained by
\begin{equation}
  \label{thetaq}
\tan (2\theta_q) \ =\ -\ \frac{2m_q|\mu R_q + A^*_q|}{
\tilde{M}^2_Q - \tilde{M}^2_q  + \cos 2\beta M^2_Z 
( T^q_z - 2e_q s^2_w )}\ .
\end{equation}
The masses  for two  scalar-quark  mass eigenstates  $\tilde{q}_1$ and
$\tilde{q}_2$ are then given by
\begin{eqnarray}
  \label{Mq12}
M^2_{\tilde{q}_1 (\tilde{q}_2)} & =& \frac{1}{2}\ \bigg\{
\tilde{M}^2_Q + \tilde{M}^2_q + 2m^2_q + T^q_z \cos 2\beta M^2_Z \nonumber\\
&& -(+)\ \sqrt{ \Big[ \tilde{M}^2_Q - \tilde{M}^2_q 
+ \cos 2\beta M^2_Z ( T^q_z - 2e_q s^2_w )\, \Big]^2\, 
                            +\, 4m^2_q |\mu R_q + A^*_q|^2 }\ \bigg\}.\quad
\end{eqnarray}
It is easy to see that the scalar quarks of the first two families are
almost degenerate for large universal  soft-scalar quark masses,  {\em
  e.g.}, bigger  than 0.5 TeV.  However,  this mass  pattern is not in
general true for the third family owing to  the non-negligible top and
bottom masses.

Defining as $\varphi^0_k  =  \sqrt{2}\, (\phi^0_k - \langle   \phi^0_k
\rangle)  = H_k + iA_k$  with $k=1,2$, we are   then able to write the
trilinear couplings of the  Higgs fields to the  scalar quarks  in the
generic form
\begin{eqnarray}
  \label{Lint}
-{\cal L}_{\rm int} &=& \sum_{q=t,b}\, \Big[\,\Big( g^{L,q}_1\, \varphi^0_1\,
+\,  g^{L,q}_2\, \varphi^0_2\, \Big) \tilde{q}^*_L\tilde{q}_L\ +\   
\Big( g^{R,q}_1\, \varphi^0_1\, +\,  g^{R,q}_2\, \varphi^0_2\, \Big) 
\tilde{q}^*_R\tilde{q}_R\, \Big] \nonumber\\
&&+\, \Big[\, \Big( h^t_1\, \varphi^0_1\, +\,  h^t_2\, \varphi^0_2\, \Big) 
\tilde{t}^*_R\tilde{t}_L\ +\ \Big( h^b_1\, \varphi^0_1\, +\,  
h^b_2\, \varphi^0_2\, \Big) \tilde{b}^*_L\tilde{b}_R\ +\quad
\mbox{H.c.}\, \Big]\, ,
\end{eqnarray}
where the different coupling parameters are given by
\begin{eqnarray}
  \label{tripar}
g^{L,t}_1 &=& \frac{c_\beta M^2_Z}{v}\, \Big( \frac{1}{2}\, -\, 
e_t s^2_w\,\Big)\, ,\qquad
g^{L,t}_2 \ =\ \frac{m^2_t}{s_\beta v}\, -\, 
\frac{s_\beta M^2_Z}{v}\, \Big( \frac{1}{2}\, -\, 
                                          e_t s^2_w\,\Big)\, ,\nonumber\\
g^{R,t}_1 &=& \frac{c_\beta M^2_Z}{v}\, e_t s^2_w\,,\qquad
g^{R,t}_2 \ =\ \frac{m^2_t}{s_\beta v}\, -\, 
                        \frac{s_\beta M^2_Z}{v}\, e_t s^2_w\, ,\nonumber\\
g^{L,b}_1 &=& \frac{m^2_b}{c_\beta v}\, -\, \frac{c_\beta M^2_Z}{v}\, 
                      \Big( \frac{1}{2}\, +\, e_b s^2_w\,\Big)\, ,\qquad
g^{L,b}_2 \ =\ \frac{s_\beta M^2_Z}{v}\, \Big( \frac{1}{2}\, +\, 
                                          e_b s^2_w\,\Big)\, ,\nonumber\\
g^{R,b}_1 &=& \frac{m^2_b}{c_\beta v}\, +\, 
                             \frac{c_\beta M^2_Z}{v}\, e_b s^2_w\,,\qquad
g^{R,b}_2 \ =\ -\, \frac{s_\beta M^2_Z}{v}\, e_b s^2_w\, ,\\
h^t_1 &=& \frac{m_t}{s_\beta v}\, \mu^* e^{-i\delta_t}\, , \quad
h^t_2 \ =\  \frac{m_t}{s_\beta v}\, A_t e^{-i\delta_t}\, , \quad
h^b_1 \ =\  \frac{m_b}{c_\beta v}\, A^*_b e^{-i\delta_b}\, ,\quad
h^b_2 \ =\ \frac{m_b}{c_\beta v}\, \mu e^{-i\delta_b}\, .\nonumber
\end{eqnarray}
{}From Eqs.\ (\ref{Lint}) and (\ref{tripar}), it is then not difficult
to compute the tree-level  couplings $H_1 \tilde{q}^*_i  \tilde{q}_j$,
$H_2 \tilde{q}^*_i  \tilde{q}_j$,  $A \tilde{q}^*_i  \tilde{q}_j$  and
$G^0  \tilde{q}^*_i  \tilde{q}_j$ ($i,j=1,2$).   These   couplings may
respectively be given by the following matrices:
\begin{eqnarray}
  \label{GHkqq}
\Gamma_0^{H_k\tilde{q}^*\tilde{q}} \!\!&=&\!\! 
\left( \begin{array}{lr} 
s_{2q}\Re e h^q_k - c^2_q g^{L,q}_k - s^2_q g^{R,q}_k & \hspace{-66pt}
-c_{2q} \Re e h^q_k - s_qc_q (g^{L,q}_k - g^{R,q}_k) + 2iT^q_z\Im m h^q_k\\
-c_{2q} \Re e h^q_k - s_qc_q (g^{L,q}_k - g^{R,q}_k) - 2iT^q_z\Im m h^q_k &
-s_{2q}\Re e h^q_k - s^2_q g^{L,q}_k - c^2_q g^{R,q}_k \end{array}
\right) ,\nonumber\\
  \label{GAqq}
\Gamma_0^{A\tilde{q}^*\tilde{q}} \!\!&=&\!\! \frac{i}{s_\beta}\,
\left( \begin{array}{cc} -is_{2q}\Im m h^q_1 & \hspace{-40pt} 
- T^q_z \Big(2\Re e h^q_1 -  \frac{\displaystyle  c_\beta s_{2q} 
                              \Delta M^2_{\tilde{q}}}{\displaystyle v} 
 \Big) + ic_{2q} \Im m h^q_1 \\  
T^q_z \Big( 2\Re e h^q_1 - \frac{\displaystyle  c_\beta s_{2q} 
\Delta M^2_{\tilde{q}}}{\displaystyle v} \Big) +
                                                   ic_{2q} \Im m h^q_1
& is_{2q}\Im m h^q_1 \end{array} \right) ,\nonumber\\
  \label{GG0qq}
\Gamma_0^{G^0\tilde{q}^*\tilde{q}} \!\!&=&\!\! 
T^q_z\ \frac{is_{2q} \Delta M^2_{\tilde{q}}}{v}\
\left( \begin{array}{cc} 0 & 1 \\
- 1 & 0 \end{array} \right)\, ,
\end{eqnarray}
where  $k  = 1,2$ and   $\Delta M^2_{\tilde{q}} =  M^2_{\tilde{q}_2} -
M^2_{\tilde{q}_1}$. Note that $\Im m h^q_1  = -\tan\beta\ \Im m h^q_2$
and $\Re  e  h^q_1  = -\tan\beta\  \Re  e  h^q_2  + (s_q  c_q   \Delta
M^2_{\tilde{q}}/v)$ in the weak basis we are working.

As has been discussed above, the tadpole parameter $T_A$ is determined
by  the $A$-boson tadpole graph,  $\Gamma^A (0)$, shown in Fig.\ 1(c).
Using the renormalization  condition  $T_A +  \Gamma^A  (0) =   0$, we
easily find that
\begin{eqnarray}
\label{TA}
T_A & = & -\,  \sum_{q=t,b}\, N^q_c\,
\int\frac{d^n k}{(2\pi)^n i}\, {\rm Tr}\, \Big[\, 
i\Delta^{\tilde{q}}(k)\, i\Gamma_0^{A\tilde{q}^*\tilde{q}}\, \Big]\nonumber\\
&= & -\, \frac{1}{16\pi^2}\, 
\sum_{q=t,b}\, N^q_c\ \frac{s_{2q}}{s_\beta}\ \Im mh^q_1\ \Delta
M^2_{\tilde{q}}\ B_0 (0,M^2_{\tilde{q}_1},M^2_{\tilde{q}_2})\, .
\end{eqnarray} 
In Eq.\   (\ref{TA}), $N^q_c =  3$  is the colour   factor for quarks,
$\Delta^{\tilde{q}}(k) =  {\rm diag} [(k^2  - M^2_{\tilde{q}_1})^{-1},
(k^2 - M^2_{\tilde{q}_2})^{-1}]$ is the scalar-quark propagator matrix
and $B_0(s,m^2_1,m^2_2)$   is  the known  Veltman-Passarino   function
defined as
\begin{eqnarray}
   \label{B0}
B_0(s,m^2_1,m^2_2) &=& C_{\rm UV}\, -\, \ln (m_1m_2)\, +\, 2\, +\, 
\frac{1}{s}\, \Big[(m^2_2-m^2_1)\, \ln\Big(\,\frac{m_1}{m_2}\,\Big)\nonumber\\
&&+\, \lambda^{1/2}(s,m^2_1,m^2_2)\,\, {\rm cosh}^{-1}
\Big(\, \frac{m^2_1+m^2_2-s}{2m_1m_2}\, \Big)\, \Big]\, ,
\end{eqnarray}
where $\lambda(x,y,z) =(x-y-z)^2-4yz$  and $C_{\rm UV}=1/\varepsilon -
\gamma_E + \ln  (4\pi\mu^2)$ is an UV  infinite constant.  For  $s=0$,
the one-loop function in Eq.\ (\ref{B0}) takes on the simple form
\begin{equation}
   \label{B00}
B_0(0,m^2_1,m^2_2)\ =\ C_{\rm UV}\, -\, \ln (m_1 m_2)\, +\, 1 +\,
\frac{m^2_1 + m^2_2}{m^2_1 - m^2_2}\, \ln\Big(\frac{m_2}{m_1}\Big)\, .
\end{equation}
A straightforward    calculation of the individual    contributions to
$H_iA$ self-energies (with $i = 1, 2$),  shown in Figs.\ 1(a) and (b),
yields
\begin{eqnarray}
  \label{SelfAH}
\Pi^{H_iA}_{(a)}(s) &=& \sum_{q=t,b}\, N^q_c\, \int\frac{d^n
   k}{(2\pi)^n i}\, {\rm Tr}\, \Big[\, i\Delta^{\tilde{q}}(k)\, 
   i\Gamma_0^{H_i\tilde{q}^*\tilde{q}}\, i\Delta^{\tilde{q}}(k+p)\, 
   i\Gamma_0^{A\tilde{q}^*\tilde{q}}\, \Big]\nonumber\\
&=&\frac{1}{16\pi^2}\, \sum_{q=t,b}\, N^q_c\ \Im m h^q_1\ 
   \Big\{\ \frac{r_i s_{2q}}{s_\beta v}\ \Delta M^2_{\tilde{q}}\ 
   B_0 (s,M^2_{\tilde{q}_1},M^2_{\tilde{q}_2})\nonumber\\
&& +\, \Re e h^q_i\ \frac{s^2_{2q}}{s_\beta}\ 
   \Big[\, B_0 (s,M^2_{\tilde{q}_1},M^2_{\tilde{q}_1}) + 
   B_0 (s,M^2_{\tilde{q}_2},M^2_{\tilde{q}_2}) -
   2B_0 (s,M^2_{\tilde{q}_1},M^2_{\tilde{q}_2})\, \Big] \nonumber\\ 
&& +\, \frac{s_{2q}}{s_\beta}\ \Big[\, (c^2_q g^{L,q}_i \, +\, 
   s^2_q g^{R,q}_i )\, \Big(B_0 (s,M^2_{\tilde{q}_1},M^2_{\tilde{q}_2}) -
   B_0 (s,M^2_{\tilde{q}_1},M^2_{\tilde{q}_1})\Big)\nonumber\\
&& +\, (s^2_q g^{L,q}_i \, +\, c^2_q g^{R,q}_i )\, 
   \Big(B_0 (s,M^2_{\tilde{q}_2},M^2_{\tilde{q}_2}) -
   B_0 (s,M^2_{\tilde{q}_1},M^2_{\tilde{q}_2})\Big)\, \Big]\ \Big\}\, ,\\  
  \label{AHtad}
\Pi^{H_iA}_{(b)}(s) &=& -({\cal M}^2_{SP})_{i2}\ =\ 
                                             \frac{r_i T_A}{v}\nonumber\\
   &=& -\, \frac{1}{16\pi^2}\, \sum_{q=t,b}\, N^q_c\ 
   \Im m h^q_1\ \frac{r_i s_{2q}}{s_\beta v}\ \Delta M^2_{\tilde{q}}\ 
   B_0 (0,M^2_{\tilde{q}_1},M^2_{\tilde{q}_2})\, ,  
\end{eqnarray}
where $s = p^2$, $r_1 = \cos\beta$  and $r_2 =  \sin\beta$. It is easy
to verify that the $H_i A$ self-energies are UV finite, only after the
CP-odd tadpole contribution given in Eq.\  (\ref{AHtad}) is taken into
account  \cite{CPodd}.  The   CP-violating  $hA$ and $HA$  self-energy
transitions are then obtained by
\begin{equation}
  \label{hHA}
\left( \begin{array}{c} \Pi^{hA}(s) \\ \Pi^{HA}(s) \end{array} \right)\ =\ 
\left( \begin{array}{c} 
\cos\alpha\ \Pi^{H_1A}(s)\ +\ \sin\alpha\ \Pi^{H_2A}(s)\\ 
-\sin\alpha\ \Pi^{H_1A}(s)\ +\ \cos\alpha\ \Pi^{H_2A}(s) \end{array}\right)\, .
\end{equation}
{}For   the  kinematic range  $M_A >    2M_Z$   and $\tan\beta  \ge 2$
($\tan\alpha  \approx   \tan\beta$),   we have   $\Pi^{hA}(s)  \approx
\Pi^{H_2A}(s)$   and    $\Pi^{HA}(s)   \approx  -   \Pi^{H_1A}(s)$.    
Furthermore, the rotation angles $\theta_t$  and $\theta_b$ are almost
equal  to  $\pi/4$,    {\em i.e.},   $s_{2q}\approx  1$,   since   the
off-diagonal elements of the scalar-quark mass matrix $\widetilde{\cal
  M}^2_q$ are much bigger than the difference of its diagonal entries.
For relatively low $\tan\beta$ values, the biggest contribution occurs
when the mass  difference between  $\tilde{t}_1$ and $\tilde{t}_2$  is
rather large. Specifically, for $2\le \tan\beta < 5$, we have
\begin{eqnarray}
  \label{hA}
\Pi^{hA}(0) &\approx& \frac{3}{8\pi^2}\, \Im m h^t_1\ \Big[\ 
\Re e h^t_2 \,\Big(\, \frac{x^2_t + 1}{x^2_t - 1}\
\ln x_t\ -\
    1\,\Big)\ -\ \frac{m^2_t}{v}\, 
\ln x_t\ \Big]\, ,\\
  \label{HA}
\Pi^{HA}(0) &\approx& -\, \frac{3}{8\pi^2}\, \Im m h^t_1 \Re e h^t_1
\,\Big(\, \frac{x^2_t + 1}{x^2_t - 1}\ \ln x_t\ 
-\ 1\,\Big)\, ,  
\end{eqnarray}
with $x_t  =  M_{\tilde{t}_2}/M_{\tilde{t}_1}    \ge 1$.    In   Eqs.\ 
(\ref{hA})  and (\ref{HA}), the CP-violating   quantities $\Im m h^t_1
\Re e h^t_1$ and $\Im m h^t_1 \Re e h^t_2$ satisfy the inequalities
\begin{equation}
  \label{ImRe}
|\Im m h^t_1 \Re e h^t_1|\ \le\ \frac{1}{2}\,\frac{m^2_t}{v^2 s^2_\beta}\ 
|\mu|^2\,,\qquad
|\Im m h^t_1 \Re e h^t_2|\ \le \ \frac{1}{2}\, \frac{m^2_t}{v^2 s^2_\beta}\
|\mu A_t|\, .  
\end{equation}
{}For  larger $\tan\beta$ values,  scalar-pseudoscalar mixings receive
significant contributions from  scalar-bottom quarks  as well, leading
to a more involved kinematic dependence.

We   shall now examine  numerically the  dependence   of $HA$ and $hA$
mixings on  the various kinematic   parameters.  Motivated by  minimal
supergravity  models, we consider the    soft scalar-quark masses  and
trilinear couplings   to be  universal: $\tilde{M}_Q =   \tilde{M}_t =
\tilde{M}_b  = M_0  =  0.5$ TeV and  $A_t  = A_b =  A$  with arg$(A) =
90^\circ$ \cite{KG}.  The soft parameters  $m^2_1$ and $m^2_2$ in  the
Higgs  potential ({\em c.f.}\   Eq.\ (\ref{LVpar})) are  also equal to
$M^2_0$ at $M_X$.  However, their renormalization-group-equation (RGE)
runnings  from  $M_X$ down  to $M_Z$  are  very sensitive   to further
model-dependent details  of  the physics  near   the unification scale
\cite{KG}.  Therefore,   we assume that $m^2_1$  and  $m^2_2$ are free
parameters  not  larger though than  few  TeV, being  subject into the
tadpole constraints in Eqs.\ (\ref{TH1})   and (\ref{TH2}).  For   our
numerical analysis, we  consider the  parameters $\{  \tan\beta,  \mu,
M_0, A  \}$ as independent and take  $M_A = M_0$.   We expect that the
above considerations  will be in good agreement  with a more elaborate
treatment.  Refinements due to the  RGE running of the MSSM parameters
are beyond the  scope  of  the   present paper   and may  be   studied
elsewhere.

In    Fig.\  2,  we display  the   dependence  of  the  $HA$  and $hA$
self-energies  on the parameter $\mu$  at vanishing momentum transfer,
$s=0$, for  three  selected values  of $|A|$.   With $\mu$ increasing,
both  the  CP-violating quantities  in     Eq.\ (\ref{ImRe}) and   the
scalar-top mass ratio $x_t$ increase and so the $HA$ and $hA$ mixings.
Indeed, we  observe  that  the  mass difference   $M_H - M_A   \approx
\Pi^{HA}(0)/   M_A$ can be as large   as  $25\% \times  M_A$, for $\mu
\stackrel{\displaystyle >}{\sim}   2$     TeV  ($M_A=0.5$ TeV)     and
$M_{\tilde{t}_1}, M_{\tilde{b}_1} >  M_Z$.  This should be  contrasted
with the  tree-level  prediction of $1.07\%  \times  M_A$ for the same
value of $\tan\beta = 2$.  As has been discussed in \cite{CPodd,APNP},
an  $HA$ mass  splitting of  order  10$\%$  of the  $A$-boson mass  is
sufficient to  give rise to significant   contributions to the neutron
EDM  through the  two-loop Barr-Zee  mechanism  \cite{BZ} close to the
observable  limit  \cite{Hayashi}.     Furthermore,  we  find     that
$\Pi^{hA}(0)$ may  reach the  $(100\ {\rm GeV})^2$  level  in the same
kinematic region and so be comparable in size  with $M_h$ at one loop. 
Nevertheless, the net effect of $hA$ mixing on the $h$-boson mass will
be a modest reduction of the  $M_h$ one-loop value by $-(\Pi^{hA})^2 /
(2M^2_hM^2_A)$, ranging from $-2\%$ to  $-10\%$ ($M_h\approx 100$ GeV)
in the $A$-boson mass interval 500 -- 200 GeV.

In Fig.\ 3, we plot  the dependence of the scalar-pseudoscalar mixings
as a function  of energy ($E_{cms} = \sqrt{s}$).   We  find that $-\Re
e\Pi^{hA}(s)$ (solid line) can increase  by one order of magnitude for
$s \approx M^2_A$ in comparison  to the $s  = 0$ value.  The cause for
this strong energy  dependence appears to  be the subtle cancellations
occurring at low energies between terms  proportional to $\Re e h^t_2$
and  $m^2_t/v$ in Eq.\ (\ref{hA})  which  get less important at higher
energies.  For similar reasons, $\Re e \Pi^{HA}(s)$ (solid line) shows
up a   milder   energy dependence because   of the   absence of   such
destructive  terms  (see also Eq.\ (\ref{HA})).    As can be seen from
Fig.\  3 too, the  absorptive  parts $\Im  m  \Pi^{HA}(s)$ and $\Im  m
\Pi^{hA}(s)$   (dashed     lines)  become     significant   above  the
$\tilde{t}^*_1 \tilde{t}_1$ threshold.  In general,  we have that both
$|\Re  e \Pi^{HA}(M^2_A)|$ and  $|\Im m \Pi^{HA}(M^2_A)|$ can formally
be of  comparable order.  Thus,  the necessary conditions for resonant
CP  violation  through $HA$ mixing   \cite{APRL} at future high-energy
machines,  such  as the  LHC,  $\mu^+\mu^-$ collider, {\em  etc.}, may
comfortably be    satisfied \cite{APNP}.     In  particular,  at     a
$\mu^+\mu^-$ collider with an  integrated luminosity of  50 fb$^{-1}$,
one may be sensitive  to $HA$ mixings up  to order  $10^{-3}$ TeV$^2$,
corresponding  to mass differences  $M_H - M_A \stackrel{\displaystyle
  >}{\sim}   1\%\times M_A$.  This  is  in  agreement  with an earlier
observation made in \cite{APRL} that CP asymmetries of order unity due
to resonant Higgs scalar-pseudoscalar  transitions can naturally occur
{\em  even} within the  MSSM \cite{BKRW}.   Finally, the dependence of
$HA$ and $hA$ mixings on $\tan\beta$ is presented  in Fig.\ 4.  We see
that    $\Pi^{HA}(0)$ decreases in   general    for  large values   of
$\tan\beta$, while $-\Pi^{hA}(0)$  exhibits  a significant enhancement
for  $\tan\beta  \stackrel{\displaystyle  >}{\sim}   50$ close to  the
perturbative bound of the theory.  The  reason for this enhancement is
due to the  sizable left-right mixing  of the scalar  bottom quarks in
the large $\tan\beta$ domain.

The fact that scalar-pseudoscalar mixings may be large within the MSSM
should not be very surprising and may be understood in simple terms as
follows.  All quartic couplings of the MSSM Higgs potential as well as
$M_h$ are suppressed by gauge coupling constants as  a result of SUSY.
On  the other   hand,  despite the   typical  loop  suppression factor
$(4\pi)^{-2}$, Higgs  scalar-pseudoscalar mixings  are enhanced by the
large top   Yukawa couplings proportional to  $m_t   \mu/v^2$ and $m_t
A_t/v^2$.  In this case it is therefore evident that naive dimensional
analysis of counting  loop suppression factors  can lead to a dramatic
underestimation of the  actual size of  the  quantum effects.   As has
been  shown by the  present  paper, a large Higgs  scalar-pseudoscalar
mixing  is indeed possible within the  MSSM and can lead to observable
CP-violating phenomena in near future experiments.

\newpage

\centerline{\Large{\bf Figure captions}}
\vspace{-0.2cm}
\newcounter{fig}
\begin{list}{\rm {\bf Fig. \arabic{fig}: }}{\usecounter{fig}
\labelwidth1.6cm \leftmargin2.5cm \labelsep0.4cm \itemsep0ex plus0.2ex }

\item[{\bf Fig.\ 1:}] Feynman graphs contributing to the $H_1A$ and 
$H_2A$ mixings: (a) One-loop self-energy graph, (b) CP-odd tadpole
renormalization, (c) Tadpole graph of the $A$ boson.

\item[{\bf Fig.\ 2:}] $HA$ and $hA$ self-energies at $s=0$ as a
  function of the parameter $\mu$.
  
\item[{\bf    Fig.\ 3:}] $HA$   and $hA$   self-energies versus energy
  $E_{cms} = \sqrt{s}$. Solid lines  indicate dispersive parts of  the
  scalar-pseudoscalar  self-energies, while dashed lines correspond to
  their absorptive parts

\item[{\bf Fig.\ 4:}] $HA$ and $hA$ self-energies at $s=0$ as a
  function of $\tan\beta$.

\end{list}

\vspace{4cm}


\begin{center}
\begin{picture}(400,200)(0,0)
\SetWidth{0.8}


\DashArrowLine(0,100)(30,100){4}\Text(10,110)[]{$A$}
\DashArrowLine(100,100)(130,100){4}\Text(130,110)[]{$H_1,H_2$}
\DashArrowArc(65,100)(35,0,180){5}\Text(65,147)[]{$\tilde{q}_1,\tilde{q}_2,
\tilde{q}_1,\tilde{q}^*_1$}
\DashArrowArc(65,100)(35,180,360){5}\Text(65,53)[]{$\tilde{q}_1,\tilde{q}_2,
\tilde{q}_2,\tilde{q}^*_2$}

\Text(65,30)[]{\bf (a)}


\DashArrowLine(180,100)(220,100){4}\Text(190,110)[]{$A$}
\DashArrowLine(220,100)(260,100){4}\Text(260,110)[]{$H_1,H_2$}
\Text(220,100.5)[]{\boldmath $\times$}\Text(220,110)[]{$T_A$}

\Text(220,30)[]{\bf (b)}


\DashArrowArc(340,120)(30,0,360){5}\Text(340,160)[]{$\tilde{q}_1,\tilde{q}_2$}
\DashArrowLine(340,50)(340,90){4}\Text(345,60)[l]{$A$}

\Text(340,30)[]{\bf (c)}

\Text(200,0)[t]{\bf Fig.\ 1}

\end{picture}
\end{center}

\begin{figure}[ht]
   \leavevmode
 \begin{center}
   \epsfxsize=17.3cm
   \epsffile[0 0 539 652]{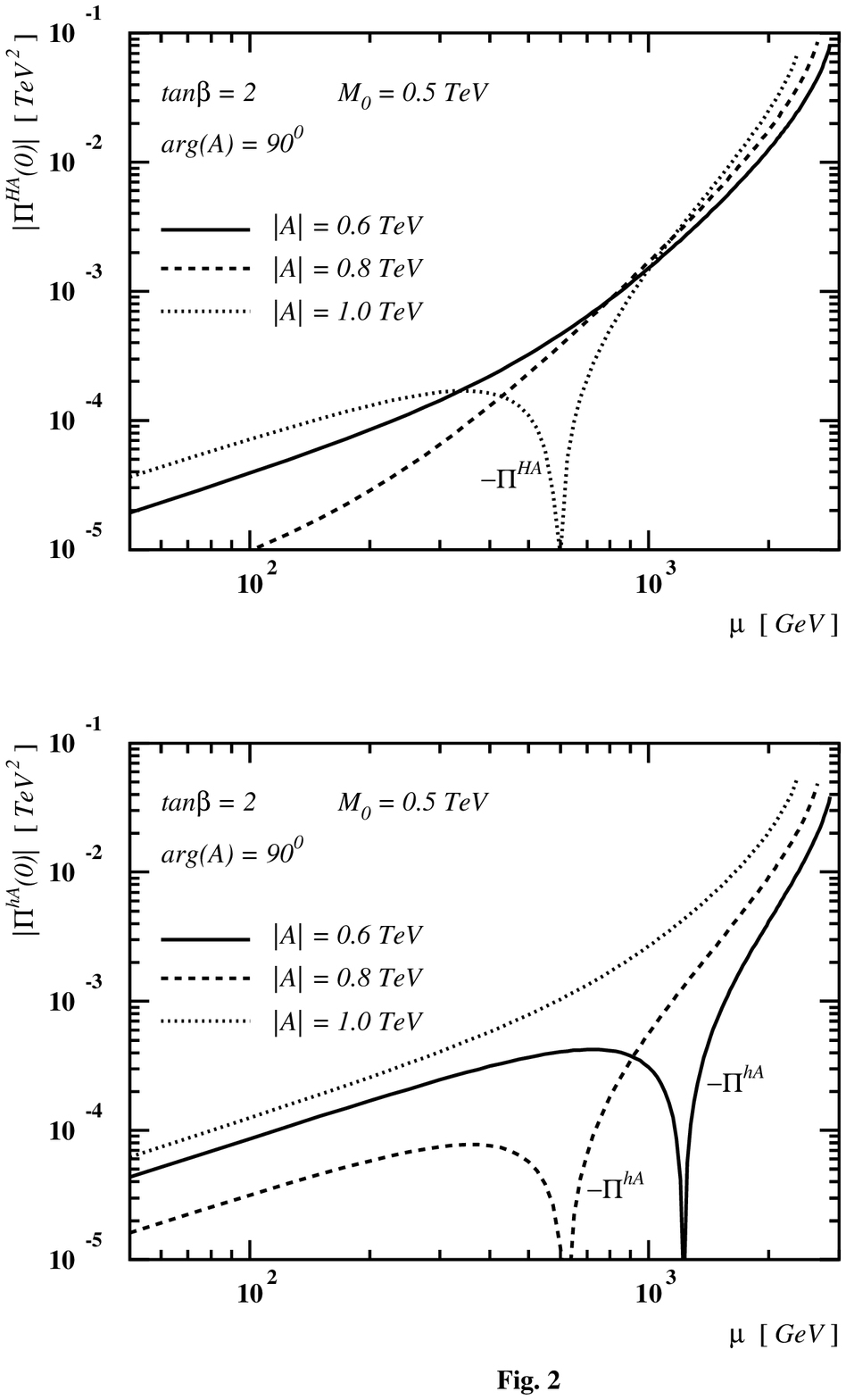}
 \end{center}
\end{figure}

\begin{figure}[ht]
   \leavevmode
 \begin{center}
   \epsfxsize=17.3cm
   \epsffile[0 0 539 652]{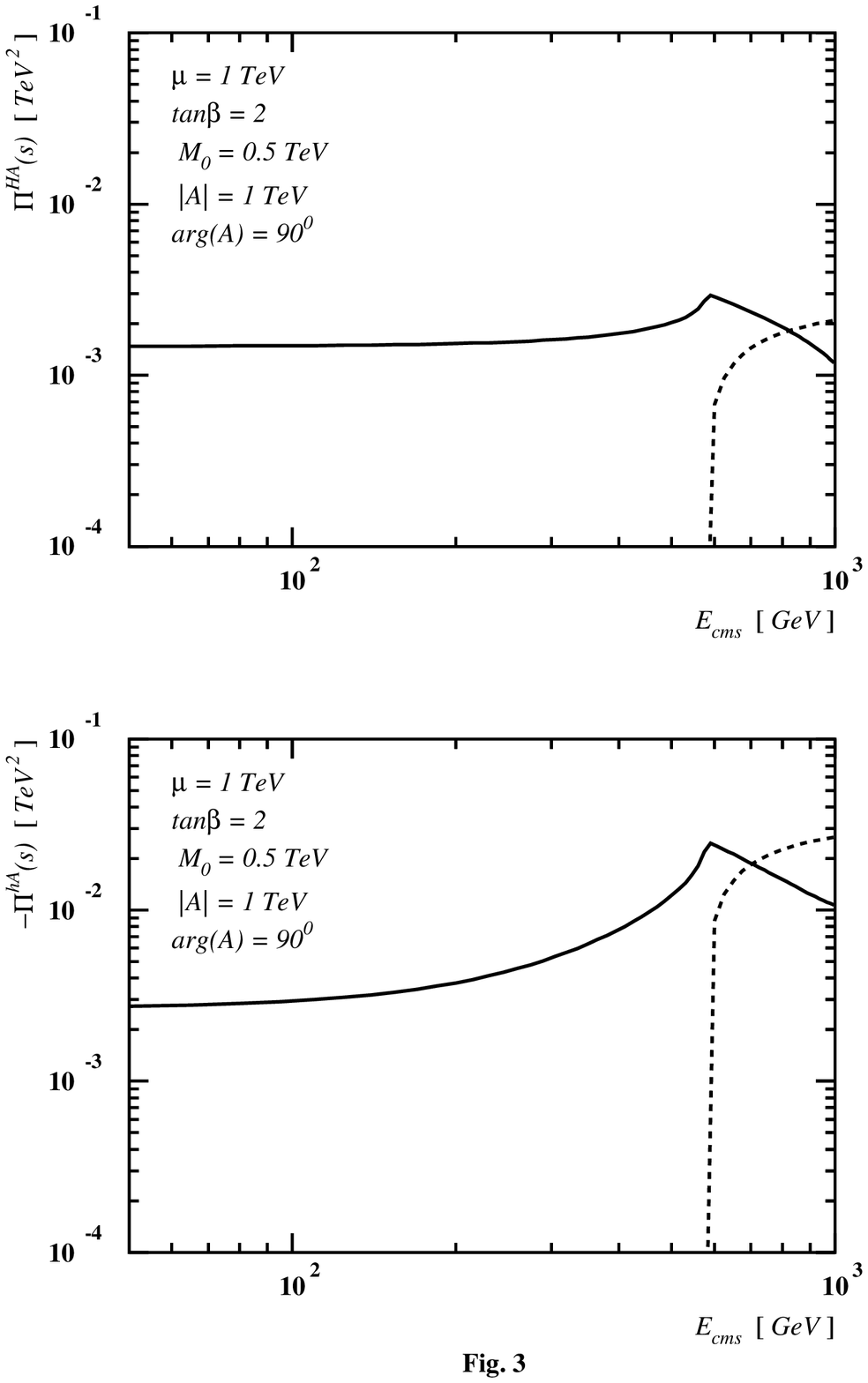}
 \end{center}
\end{figure}

\begin{figure}[ht]
   \leavevmode
 \begin{center}
   \epsfxsize=17.3cm
   \epsffile[0 0 539 652]{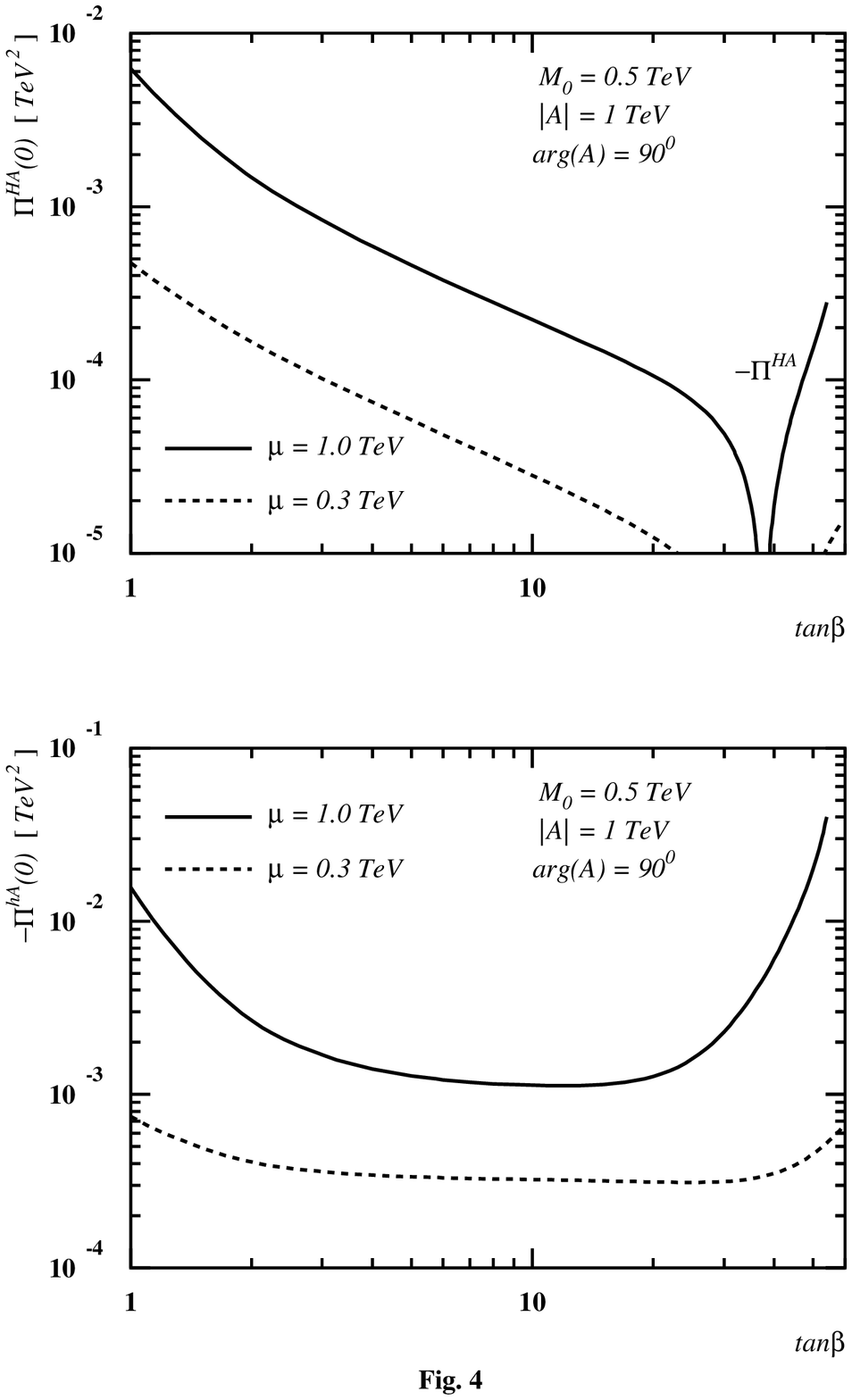}
 \end{center}
\end{figure}

\end{document}